\documentclass[reqno]{amsart} 
\usepackage{amssymb,amsmath,amsfonts,amsthm,comment,mathrsfs,times,graphicx}
\usepackage[T1]{fontenc}
\usepackage{lmodern}

\usepackage{geometry}                
\title[A mesoscopic theory ...phase transitions with nonlinear elasticity]{A mesoscopic theory to describe the flexibility regulation in F-actin networks: An approach of phase transitions with nonlinear elasticity}

\author{Horacio Lopez-Menendez}
\thanks{Institut Jacques Monod (IJM), CNRS UMR 7592 et Universit\'e Paris Diderot, 75013 Paris, France}
\thanks{To whom correspondence should be addressed. E-mail: horacio.lopez.menendez$@$gmail.com}

\begin{document}



\begin{abstract}
The synthetic actin network arouses great interest as bio-material due to its soft and wet nature that mimics many biological scaffolding structures. Inside the cell, the actin network is regulated by tens of actin-binding proteins (ABP's), which make for a highly complex system with several emergent behaviours. In particular, calponin is an ABP that was identified as an actin stabiliser, but whose mechanism is still poorly understood. Recent experiments using an in vitro model system of cross-linked actin with calponin and large deformation bulk rheology, found that networks with basic calponin exhibited a delayed onset and were able to withstand a higher maximal strain before softening. 

In this work, we show that the difference between the two networks, with and without calponin, not only provides bundle flexibility. But, also encodes alterations in the pre-strain and the regulation of the crosslinks adhesion energy to define the new yielding point. We verify these effects theoretically using nonlinear continuum mechanics for the semiflexible and crosslinked network. In addition, the alterations over the microstructure are described by the definition of an interaction parameter $\Gamma$ according the formalism of Landau. According to this simple model we demonstrates that the interaction parameter can describe the experimental observations following an scaling exponent as $\Gamma \sim \mid c-c_{cr} \mid^{1/2}$, where $c$ is the ratio between concentration of calponin and actin. This result provides interesting feedback to improve our understanding of several mechano-biological pathways.
  
\end{abstract}

\maketitle

\section*{Introduction}

The actin network is one of the most relevant structural components inside cytoskeleton of eukaryotic cells, and has a highly dynamic mechanical properties in the cell. Which is essential to many processes, such as cell adhesion, mechanosensing and mechanotransduction, motility, and cell shape among others. At the same time, the development of actin based biomimetic systems at micro and nano-scale, demand a deep understanding of their mechanical properties at large deformations, where the non-linear mechanics encodes effects that allow the tuning of unexpected effects \cite{Schmoller2010,Schmoller2013,lopez2016}.

Significant progress in the past decades has provided a rather complete picture of the linear mechanical response to small applied stresses or strains \cite{rigato2017}. However, cells are often subject to large deformations and reach nonlinear regimes that are far from being well understood \cite{lieleg2010,koll2011,broedersz2014,Schmoller2010}. The fascinating mechanical properties of living cells are mediated by the cytoskeleton (CSK), a dynamic network of filamentous proteins composed of actin filaments, microtubules, and intermediate filaments \cite{gardel2008,koll2011,fletcher2010,blanchoin2014}. This complex structure has solid and liquid-like behaviours. The solid one is associated with strongly cross-linked actin filaments, which resist sliding and accumulate tension (fimbrin and fascin are compact cross-linking proteins that create parallel-aligned actin networks or bundles, they are found in stiffer protrusions as filopodia \cite{dos2003,tseng2005}. On the other hand the liquid-like behaviour, is associated with weakly and transient cross-linking proteins, such as $\alpha$-actinin or filamin, which produce networks, and slide more readily, enabling to soften and, or flow as a liquid. 

The ability of the network with calponin to resist large deformations without breaking, is highly relevant for several biological mechanisms. The toughening mechanism in living structures is a complex process with many unresolved issues. Recently it was described that the effects of super-elasticity into the cytoskeleton is a relevant topic to allow the large deformations during the formation of domes \cite{latorre2018}. In addition, into the context of hydrogels, the way in which the structure can resist the fracture without collapsing is by doing interpenetrating polymer networks, made by two or more polymeric networks, at least one of which is polymerised and/or crosslinked in the presence of the other. The polymeric networks are interlaced on a molecular scale but not covalently bonded to each other \cite{sun2012,nakajima2013, zhao2014, ducrot2014}. In this sense the improvement of the understanding of the micromechanics of these gels through, a constitutive modelling of interpenetrating networks, has also been proposed \cite{suo2009,zhao2012}. 

However, concerning the cytoskeleton, the process seems to be more complex. The dynamic of polymerisation and depolymerisation of the F-actin usually allows the self-healing and the crosslinks are able to rebind under certain conditions, in addition to that, the regulation of the dynamics of the actin filaments is linked with a wide variety of actin binding proteins (ABPs) \cite{gardel2008, fletcher2010}. Specifically the mechanics of the actin cytoskeleton is regulated by upward sixty known actin-binding proteins (ABPs) defining different emergent behaviours. They have the ability to control their fraction of passive and active ABPs, to regulate their own rheology. One specific protein that plays a highly relevant role is the calponin ABP. It was discovered in smooth muscle cells and was studied as a possible regulator of actomyosin interaction \cite{winder1990}. Moreover, in non-muscle cells, calponin is known to be involved in actin stabilisation of stress-fibres and to increase the tensile strength of the tissue under strain, among other effects \cite{jensen2014}. However, the growing evidence on the role performed by calponin as a structural stabiliser, the underlying micro-mechanic is still unknown.

In order to treat the mechanical effect developed by calponin inside the actin structure \cite{jensen2014}, studied an in-vitro F-actin network crosslinked with calponin to gain insights about its mechanical properties using large deformation rheology. They found that the networks with calponin are able to reach a higher failure stress and strain while reducing the pre-strain of the network. Calponin delays the onset of network stiffening, something observed in polymer networks with increased flexibility. They also observed that the micro-structural origin of this behaviour was related to the decrease on the persistence length at a single filament level. 

In order to address theoretical and computational models for biopolymeric networks many approaches have been developed, providing new ways of thinking about the cells and tissues mechanics. On the one hand microstructural approaches based on the worm-like chain model represent good a description of the actin mechanics \cite{broedersz2014, lopez2016, meng2017, vernerey2018, ferreira2018}. On the other hand computational large-scale models have made a relevant progress to address complex dynamics between filaments and crosslinks \cite{Kim2009b,zagar2011,borau2012,vzagar2015,tito2019h}.

Therefore, to explain the observed effect reported by \cite{jensen2014} we propose a new mathematical model into the framework of non-linear continuum mechanics by using the semiflexible filament described by a worm-like chain following the Blundel-Terentjev formalism \cite{blundell2009}, and to homogenise the F-actin network we follow the 3-chain model as was implemented  by Meng et al \cite{meng2016, meng2017}. On the basis of this model, we introduce the dynamic effect of the crosslinks in order to capture the strengthening-weakening transition manifested by the network \cite{lopez2016, lopez2017, lopez2019}. Next, to capture the effects of the interaction between actin and calponin, we propose an energy term associated to the interaction energy, by using the Landau model for continuous phase transition formalism. This term allows us to define the interaction parameter that captures the effects of the alteration over the F-actin network due to the interaction \cite{nishimori2010, lopez2019}. Finally we compare with experimental measurements performed by Jensen et al. \cite{jensen2014},and discuss the results and future works.

\section*{Methods}
\subsection*{Mechanics F-actin network with transient crosslinks}

First we consider the definition of the main concepts associated with non-linear elasticity used to describe the model:
Let $\Omega_{0}$ be a fixed reference configuration of the continuos body of interest (assumed to be stress free). We use the notation $\mathbf{\chi}:\Omega_{0}\rightarrow\mathbb{R^{\text{3}}}$ for the deformation, which transforms a typical material point $\mathbf{X}\in\Omega_{0}$ to a position $\mathbf{x}=\mathbf{\chi}(\mathbf{X})\in\Omega$ in the deformed configuration. Further, let  $\mathbf{F}(\mathbf{X})=\frac{\partial\mathbf{\chi}(\mathbf{X})}{\partial\mathbf{X}}$ be the deformation gradient and $J(\mathbf{X})=\det\mathbf{F}(\mathbf{X})>0$ the local volume ratio. Then, let consider a multiplicative split of $\mathbf{F}$ into spherical (dilatational) part, $\mathbf{F}_{V}=(J^{1/3}\mathbf{I})$ and a uni-modular (distortional) part $\bar{\mathbf{F}}$. Note that $\det\bar{\mathbf{F}}=1$. 

We use the right and left Cauchy-Green tensors denoted by $\mathbf{C}$ and $\mathbf{b}$, respectively, and their modified counterparts associated with $\bar{\mathbf{F}}$, $\bar{\mathbf{C}}$ and $\bar{\mathbf{b}}$, respectively. Hence,
\begin{equation}
\mathbf{C}=\mathbf{F}^{T}\mathbf{F}=J^{2/3}\bar{\mathbf{C}},
\end{equation}
\begin{equation}
\bar{\mathbf{C}}=\mathbf{\bar{F}}^{T}\mathbf{\bar{F}},
\end{equation}
\begin{equation}
\mathbf{b}=\mathbf{F}\mathbf{F}^{T}=J^{2/3}\bar{\mathbf{b}},
\end{equation}
\begin{equation}
\bar{\mathbf{b}}=\bar{\mathbf{F}}\bar{\mathbf{F}}^{T}.
\end{equation}

Now, we consider as a first approximation the Helmholtz free energy which accounts the strain energy functions associated with the elasticity of the crosslinked F-actin network with and without the decoration with he calponin ABP. In this function the mechanical strain deformation is described by the Cauchy-Green tensor $\mathbf{C}$. In addition we incorporate an energy term associated with the interaction between actin and calponin which is proportional to the ratio between the concentration of calponin and the concentration of actin $c=\frac{[calponin]}{[actin]}$. This potential energy will allow us to define an interaction parameter $(\Gamma)$
\begin{equation}
\Psi(\mathbf{C},\Gamma,c)=\Psi_{actin}(\mathbf{\bar{C}},\Gamma;c)+\Psi_{inter}(c,\Gamma)+U(J), \label{FreeE}
\end{equation}

In the following we first describe the energy functions without considering the effects of the interaction defined by $\Gamma$, which will be defined further for clarity. The first term represents the strain energy function (SEF) for the crosslinked actin network. This is modelled by means a SEF based on the wormlike chain model for semi-flexible filaments.  In order to do so, we propose a mathematical model into the framework of non-linear continuum mechanics by using the semiflexible filament described by a worm-like chain following the Blundel-Terentjev formalism \cite{blundell2009}.The two main physical parameters are the contour length of the filament, $L_c$, and the persistence length $l_p$, which represents a measure of the bundle stiffness and it compares the bending energy with the thermal energy, $l_p=EI/k_BT$. The chain is considered as semiflexible when $L_c \sim l_p$, we also  define $c_s=l_p/2L_c$ as a dimensionless stiffness parameter reflecting the competition between bending and thermal energy. Combining the effects of enthalpy arising from bending and entropy of conformational fluctuations, the closed form of the single chain free energy can be expressed as a function of its end-to-end factor, $x=\xi/L_c$: 
\begin{equation}\label{energy}
\psi_{chain}=k_BT\pi^2 c_s (1-x^2)+\frac{k_BT}{\pi c_s(1-x^2)}.
\end{equation}

Next, we build the continuum elastic free energy of the network; in this sense several ways to perform the homogenisation have been proposed into the context of rubber and biopolymer based on the eight chains model, \cite{Arruda1993,Palmer2008,lopez2017,lopez2016} or by micro-sphere integration, \cite{ferreira2018}. Here, we choose to apply the three chain scheme, as was proposed by Meng et al, \cite{meng2016} because it allows the correct calculation of the normal stress.
The primitive cube for the homogenisation is constructed with lattice points representing the crosslink sites, and the edges are aligned along the principle directions of deformation tensor $\mathbf{C}$. Three chains are linked with their end-to-end vectors along the edges and the equilibrium mesh size $\xi$. On deformation, the lengths of the perpendicular edges over the lattice point become $\lambda_1 \xi, \lambda_2\xi$ and $\lambda_3\xi$ respectively. Then the free energy density of a semiflexible network can be expressed as:
\begin{equation}
\Psi_{3c}(\lambda_{i=1,2,3})=\frac{n}{3}\sum_{i=1,2,3}\psi_{chain}(\lambda_i\xi)
\end{equation}
\begin{equation}
\Psi_{3c}=\frac{nk_BT}{3}\left[\pi^2c_s(3-I_1x^2)+\frac{3-2I_1x^2+I_2x^4}{\pi c_s(1-I_1x^2+I_2x^4-I_3x^6)}\right],
\end{equation}

with $x=\xi/L_c$

If the stress tensor  is expressed as a function of the strain invariants for an incompressible material, where the $I_3=1$, it can be expressed as:

\begin{equation}\label{s3s}
\mathbf{\sigma}=2\left[\left(\frac{\partial\Psi}{\partial \bar{I}_1}+\bar{I}_1\frac{\partial\Psi}{\partial \bar{I}_2}\right)\mathbf{\bar{C}}-\left(\bar{I}_1\frac{\partial\Psi}{\partial\bar{I}_1}+2\bar{I}_2\frac{\partial\Psi}{\partial\bar{I}_2}\right)\frac{\mathbf{I}}{3}-\frac{\partial\Psi}{\partial\bar{I}_2}\mathbf{\bar{C}.\bar{C}}\right]-p\mathbf{I}
\end{equation}

\begin{figure}[h]
	\includegraphics[width=13cm]{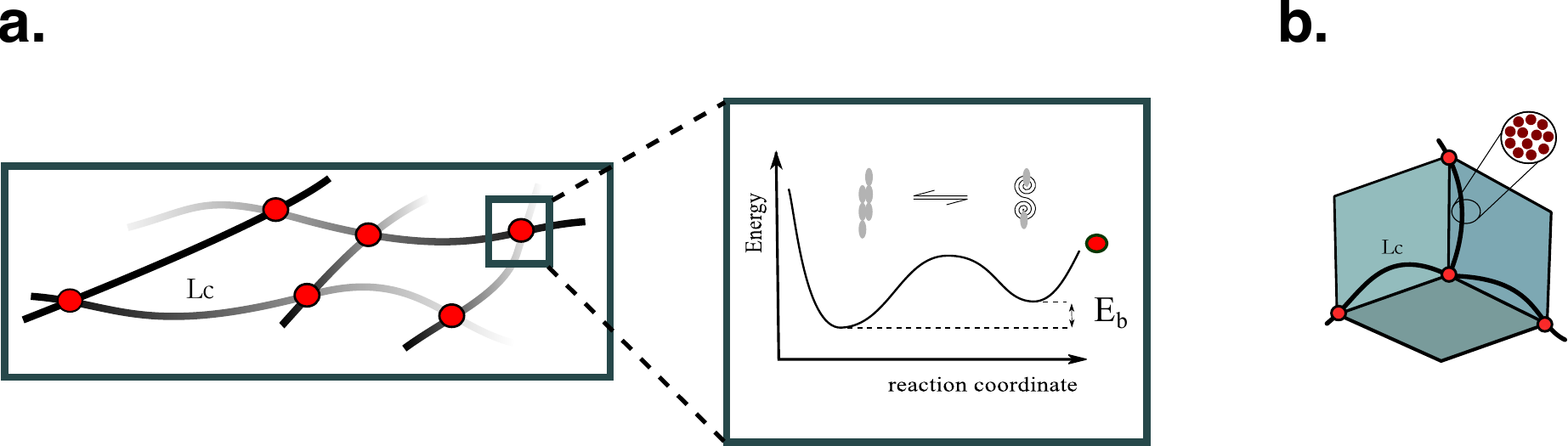}
	\centering
	\caption{(a) Three-chain homogenisation lattice. (b) filaments and crosslinks energy landscape. (c) unbinding probability $P_{ub}$. }
	\label{chainsfig}
\end{figure}

The mechanical properties of the network are defined by the sort of interaction between actin with crosslinks. 
Formerly, we have described the constitutive model for a semiflexible network with rigid crosslinks (covalent bonds). In the following, we address the necessary modifications into the model to capture the network fluidisation due to the transient chemical crosslinks.  These crosslinks are not covalent bonds (with high adhesion energy) in fact their adhesion energy is in the order of tens of $k_BT$, and have a transient dynamics \cite{ferrer2008}.
In general terms, this kind of gels with chemical crosslinks (proteins as $\alpha$-actinin) behaves as physical gels \cite{de1979}. Then, to capture this interaction and the size of the mesh, i.e the contour length, $L_c$, we propose, in a similar manner as proposed by Lopez-Menendez et al. \cite{lopez2016}, the following expression as:
\begin{equation}
L_{c}=L_{c}^{\min}+\delta L_c^{cl}P_{ub},
\end{equation}
Where the adhesion is modelled as a two states process in which $P_{ub}$ defines the unbinding probability. With that name we denote that, from a mesoscale perspective, the network can fluidise if the crosslinks undergo any transition from a bind, fold or rigid towards a state of unbind, unfold or flexible. $L_{c}^{\min}$ represents the contour length when $P_{ub}=0$ (bind crosslink), and $\delta L_c^{cl}$ represents the average increment of the contour length when the unbinding probability is one. Then, this sort of interactions can be modelled as  a reversible two-state equilibrium process \cite{Brown2009,lieleg2010,purohit2011}. Moreover, taking into account that the shear velocity is much slower than the internal crosslinks dynamics we can consider the interaction at steady state. Then it can be expressed as:
\begin{equation}
\frac { P_{ ub} }{ P_{ b} } =\exp -\frac {( E_b - w_{ ext} )}{ k_{ B }T },  \label{eq_7}
\end{equation}
Where two-state model has the bind state as the preferred low free energy equilibrium state at zero force and the unbind state as the high free energy equilibrium state at zero force.  In addition,$E_b$  represents the difference in the free-energy between these states.  $w_{ext}$ represents the external mechanical work that induces the deformation of the crosslink. In the following we write an expression for the unbinding probability considering the shear strain as the main driving force, by using scaling arguments \cite{Bell1978,de1979}. Then, in order to do so  we can re-write as  $w_{ext}=f.a$, where $a$, is the distance between states in the direction of the reaction coordinate, it is a length scale in the order of the monomer size. The force $f$ can be expressed as $f\sim G\gamma \xi^2$ in which $\gamma$ is the shear strain, $G$ is the shear modulus which can be estimated as $G\sim\frac{l_p k_BT}{L_c\xi^3}$, and $\xi$ the network mesh size. Also, taking into account that the unbinding transition due to the bundle strain happens in the semiflexible regime when $\xi \sim L_c$. Therefore, reorganising the terms we arrive to an expression for the $P_{ub}$ as a function of the shear strain as: 
\begin{equation}\label{eq_pub}
P_{ub}=\frac{1}{1+\exp\left[\kappa\left(\gamma_{0}-\gamma\right)\right]}, \quad
\gamma_0\sim\left(\frac{E_b \xi^2}{k_BT l_p a}\right)
\end{equation}
where the parameter $\kappa\sim\frac{l_p a}{\xi^2}$ is proportional to the sharpness of the transition between states and $\gamma_0$ is the characteristic strain which is proportional to the adhesion energy; it defines the point at which the probability of unbinding is 0.5. If $\gamma_{0} << \gamma$, the network is easy to be remodelled showing a fluid-like behaviour. On the contrary, if $\gamma_{0} >> \gamma$, the crosslinks are stable and the probability of transition is low, consequently the network behaves as a solid-like structure. Moreover, we can clearly identify that the characteristic strain $\gamma_0$, scales proportionally with the adhesion energy $E_b$, with the mesh size $\xi$ and increases when the bundle stiffness $l_p$, becomes smaller.


\subsection*{Interaction between actin and calponin}

In order to consider the interaction between networks we expect that for very low concentrations, the changes over the mechanical response of the actin network is almost negligible, but once a certain value is overpassed, the effects associated with the interaction are more relevant until the interaction reaches some asymptotic value. This effect can be interpreted by using arguments from phase transition. According to that, we propose to use an interaction energy $\Psi_{int}$ by means the Landau functional that couples the effects with the networks \cite{nishimori2010, lopez2019}. This energy is written in terms of an interaction parameter defined $\Gamma=\Gamma(c)$ where $c$ represents the ratio between the concentrations of vimentin and actin.  As we are interested to know when the effect of calponin becomes relevant on the mechanical response, we focus on the critical phenomena, when the concentration of $c$ is near the critical point and the interaction parameter $\Gamma$ assumes a very small value. 
This allows us to expand the free energy in even powers of $\Gamma$ and retain only the lowest order terms. 
 
Then we re-write the Helmholtz free energy as follows:
\begin{equation}
\Psi( \bar{\mathbf{C}}, \Gamma)=\frac{\alpha}{2}\Gamma^2 + \frac{\beta}{4}\Gamma^4 + \Psi_{actin}(\bar{ \mathbf{C}}, \Gamma) 
\end{equation}

\begin{figure}[h]
	\includegraphics[width=14cm]{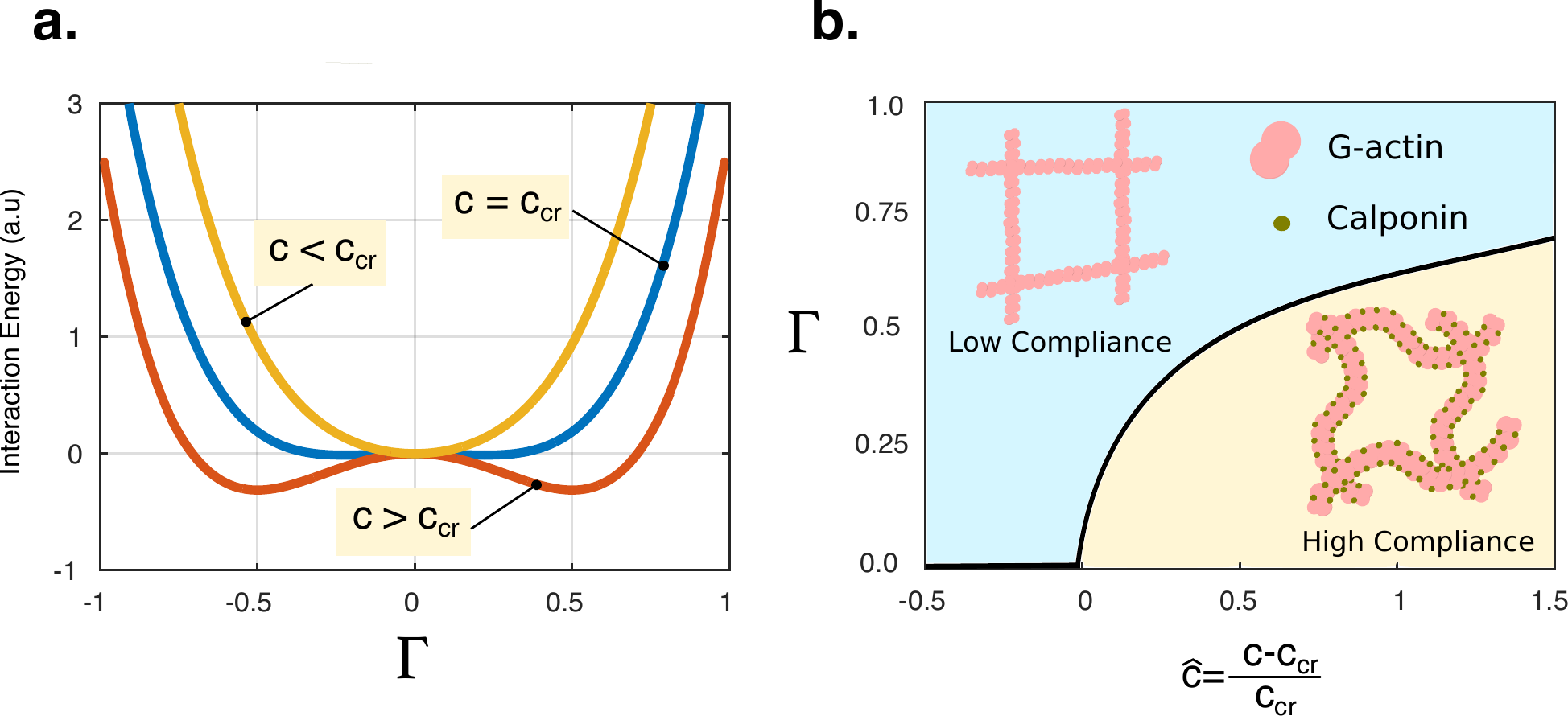}
	\centering
	\caption{\textbf{a.} Energy landscape for the interaction energy for the order parameter $\Gamma$ for concentration ratios above, exactly at, and below the transition concentration ratio for the phase transition.	\textbf{b.} Interaction parameter $\Gamma(c)$ as a function of the reduced concentration ratio $\hat{c}$, showing the continuous, second order, phase transition for the crosslinked actin network from a state of low compliance towards a state of high compliance, due to the role played by the calponin.} \label{calp_eff}
\end{figure}


Where the first two terms define the Landau energy associated with the interaction parameter; the third term interprets the strain energy for the network without any coupling as a function of the isochoric Cauchy strain tensor and the interaction parameter. The interaction energy is described as an expansion in even powers for the Landau energy term, associated with the remodelling, $\Psi_{inter}(\Gamma,c)$, and is considered the simple way to represent a two well energy potential. The behaviour from the equilibrium position (minimum) of the interaction energy changes at $\alpha=0$, we identify this point with the critical point at $c=c_{cr}$. It allows us to choose $m\hat{c}$ as $\alpha$, where $m$ is a positive constant and $\hat{c}=(c-c_{cr})/c_{cr}$ is the deviation of the concentration ratio from the critical point normalised by $c_{cr}$ which we define as a reduced concentration ratio. Then, the simplest election is $\alpha=m\hat{c}$, for which $\alpha>0$ above the critical point and $\alpha<0$ below. The dependence of $\beta$ with $\hat{c}$ does not affect qualitatively the behaviour of the free energy in the vicinity of the critical point and therefore we take $\beta$ as a constant. Then, minimising the free energy to obtain the equilibrium condition with respect to the interaction parameter $\Gamma$, it yields:
\begin{equation}
\frac{\partial \Psi}{\partial \Gamma}\approx2\alpha\Gamma+4\beta\Gamma^3=0.
\end{equation}
Thus, the equilibrium value of remodelling, $\Gamma$ is
\begin{equation}
\Gamma\approx\pm\left( \frac{-\alpha}{2\beta}\right)^{1/2} =\left\{ 
\begin{array}{rcl}
 \pm\left[\frac{m(c-c_{cr})}{2\beta c_{cr}} \right]^{1/2}  &\forall &  c>c_{cr} \\
  0 &\forall \quad & c<c_{cr}		
\end{array}
\right.
\end{equation}
The interaction parameter is canceled when the concentration ratio $c \approx c_{cr}$, above the critical value scale as $\Gamma \sim {(c-c_{cr})}^{\frac{1}{2}}$. For values of $c$ below the critical, the level of interaction is zero. 

Then, once the interaction parameter has been defined, we will describe in the following, the internal variables that encode the interplay between the two networks and how they are driven by the interaction parameter $\Gamma(c)$. We consider the following hypothesis: During the network buildup where the monomers are polymerised and the filaments crosslinked the emergent physical interaction between bundles produces the formation of trapped stress into the structure, associated with the physical interaction between filaments. This effect is compensated by the deformation of the bundle and the chemical crosslinks, and the interaction is dependent of the stiffness ($\epsilon \propto l_p$). As a consequence, it is potentially able to induce conformational changes over the crosslink structure, as was described by Golji et al.\cite{Golji2009}. This effect can be explained by the equation $\ref{eq_pub}$ for the critical strain, in which the dependence with the persistence length scales as $\gamma_0 \propto \xi^2 l_p^{-1}$.
Then, in order to describe the phenomenological phase transition by using the interaction parameters $\Gamma(c)$, the variables that encode the alterations in the rheological response are described as a linear perturbation of $l_p,\epsilon$ and $\gamma_o$ as:

\begin {equation}
        l_p(\Gamma)  = \bar{l_p} -\delta l_p \Gamma;  \quad
      \epsilon(\Gamma)  = \bar{\epsilon} -\delta \epsilon \Gamma;  \quad
        \gamma_0(\Gamma)  = \bar{\gamma_0} +\delta \gamma_0 \Gamma.
 \end{equation}

\section*{Results}
The proposed theory is used to describe the experiments conducted by Jensen et al.\cite{jensen2014} on the \textit{in-vitro} crosslinked F-actin networks, where the actin is decorated with and without calponin. We evaluate the proposed model with the set of parameters that better fit experiments of monotonic shear tests at large deformation from~ Jensen et al.\cite{jensen2014}. Taking into account that all the experiments where performed at the same velocity, where the time of the experiments is much higher that the characteristic relaxation time of the network, its response can be considered in a quasi-static regime.

Then, the coupled set of equations becomes:
%
%
%


\begin{itemize}
   
    \item \textbf{interaction actin-calponin}: $\forall c > c_{cr}:$
    	$ \begin{cases} 
	 		        l_p(c) = \bar{l_p} - \delta l_p \left[\frac { m ( c-c_{cr} ) }{2\beta c_{cr}} \right]^{1/2} , \\ 
	 		\epsilon(c) = \bar{\epsilon} - \delta \epsilon \left[\frac{m(c-c_{cr})}{2\beta c_{cr}} \right]^{1/2}, \\ 
		    \gamma_0(c) = \bar{\gamma_0} + \delta \gamma_0 \left[\frac{m(c-c_{cr})}{2\beta c_{cr}} \right]^{1/2} 
          \end{cases}$
    \item  \textbf{contour length:} 
        	\begin{equation}\label{eqLc}
		L_{c}(c,\gamma)=L_{c}^{0} + \frac{\delta L_c^{cl}}{1+\exp\left[\kappa\left(\gamma_{0}(c)-\gamma\right)\right]}. 
	\end{equation}
    \item \textbf{reference configuration:} 
    	\begin{equation} \label{refconf}
		x(c,\gamma) = \left[ 1+\epsilon(c) \right ]  \left[ 1-\frac{2L_c (c,\gamma)}{l_p (c)\pi^{3/2}} \right]^{1/2} 
	\end{equation}
    \item \textbf{shear stress:} 
         \begin{equation}
         	\sigma_{xz}=\frac{2}{3}nk_BT\gamma x(c,\gamma)^2 \left[\frac{(1-x(c,\gamma)^2)} {c_s\pi \left[1-(2+\gamma^2)x(c,\gamma)^2+x(c,\gamma)^4\right]^2}-c_s\pi^2\right]
	\end{equation}
\end{itemize}

The parameters of the model can be divided in two types. On the one hand, the rigid-wormlike chain parameters $L_{c}^{0}$, $l_{p}$, $\delta L_{c}$,  and $\epsilon$ which are of the order of magnitude of the values used to describe experiments of \textit{in-vitro} F-actin networks and to keep on the regime of semi-flexible entropic elasticity \cite{Palmer2008, meng2016,lopez2016}. On the other hand, the parameters associated with the remodeling dynamics of the crosslinks $\kappa$ and $\gamma_{0}$, and the parameters that describe the interaction parameter $\Gamma(c)$. These parameters encode the transitions to induce the fluidisation of the network and represent an indirect measure of the adhesion force of crosslinks.The associated values are: $n\sim 1e18$; $l_p \sim L_c^0$; $ \frac {\delta L_c^{cl}}{L_c^0}=1$; $\frac{\delta\epsilon}{\epsilon}=0.03 $; $\frac{\delta l_p}{\bar{l_p}}=2.48 $; $\frac{\delta\gamma_0}{\bar{\gamma_0}}=1.789$; $\kappa=10$; $\frac{m}{2\beta}=0.29$;.

 \begin{figure}[h!]
	\includegraphics[width=15cm]{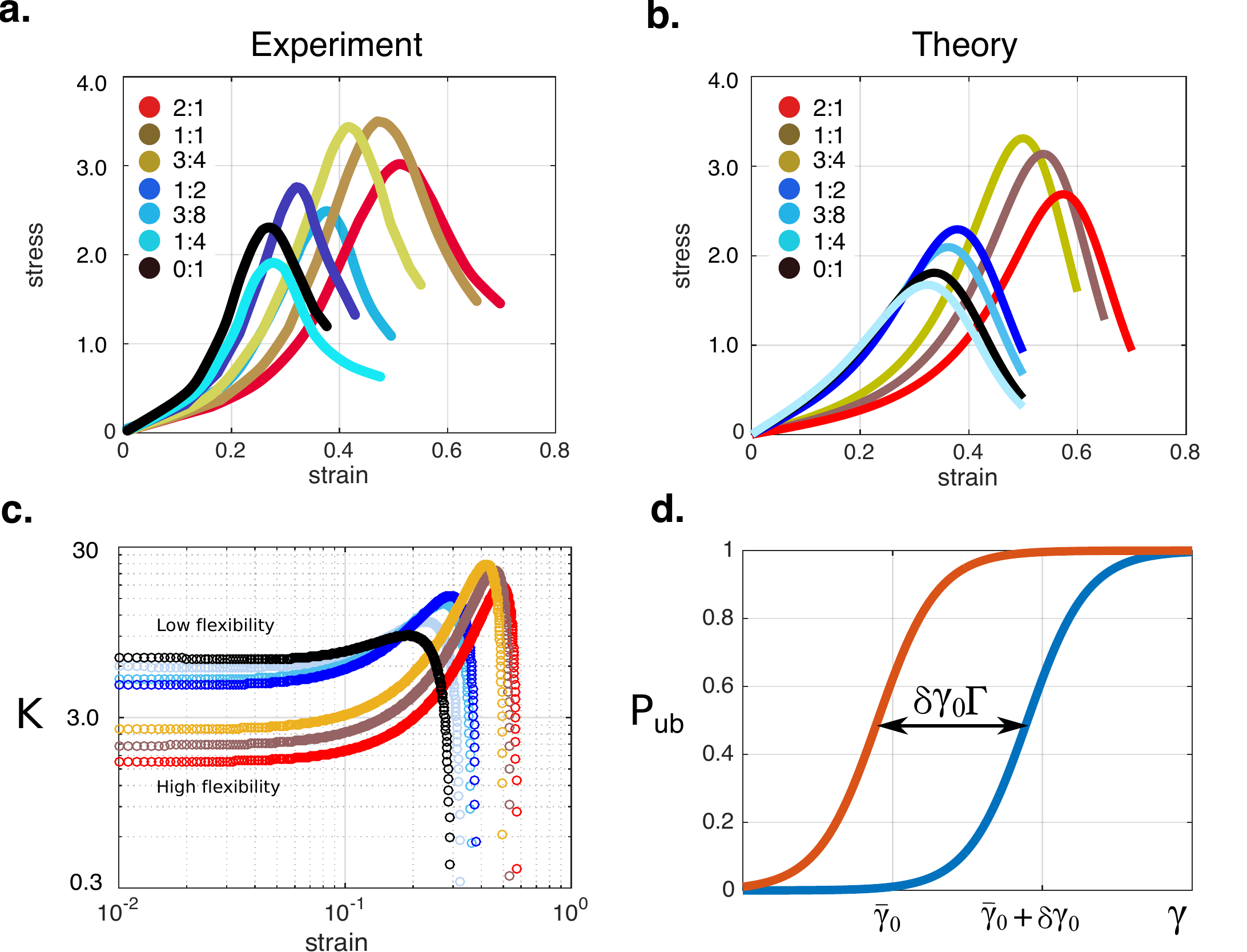}
	\centering
	\caption{
	\textbf{a.} Experimental results from bulk rheological stress-strain curves for actin networks with different ratios of $c=[calponin/actin]$. It can be observed that as much as the concentration of calponin increases, the network is able to reach higher yielding stress and strain points. \textbf{b.} The simulation simulation for the bulk rheological stress-strain curves with the model. The propossed theory is able to capture the general trend followed by the experimental curves. 
\textbf{c.} Differential elastic modulus $K\approx\frac{d\sigma}{d\gamma}$ showing the regions with high and low compliance. \textbf{d.} Unbinding probability $P_{ub}(\gamma,\Gamma)$ as a function of the shear strain $\gamma$ and the interaction parameter $\Gamma$. }\label{results}
\end{figure}

Jensen et al.\cite{jensen2014} studied bulk rheology stress-strain curves of cross-linked actin networks with calponin/actin molar ratios $c$ $(0.25; 0.375; 0.5; 0.75; 1; 2)$ and revealed a gradual shift in yield strain. These changes were evident even at sub-saturating concentrations of calponin, suggesting that actin networks partially decorated with calponin exhibit changes similar to, but less pronounced than, those seen in fully decorated networks \ref{results}.a. The model results are showed in the figure \ref{results}.b, where we can clearly see a qualitative good agreement between the theoretical model and the experimental observations, \ref{results}.a. 

In addition to that, the figure \ref{results}.c illustrates the effect of pre-tension introduced into the networks at $\gamma=0$, where the level of compliance increase with the increases of the ratio calponin copolymerised with actin. Next, another important aspect introduced by the calponin is the extension of the solid-like regime. This effect is incorporated into the model by the regulation of the characteristic strain $\gamma_0$, which is a function of the interaction parameter with the term $\delta \gamma_0 \Gamma(c)$. The figure \ref{results}.d shows the unbinding probability $Pub$ that encode the crosslinks state. Then, as much as the concentration of calponin increase, the unbinding probability decrease and extend the solid-like regime, by increasing the yielding point of stress and strain.

 \begin{figure}[h!]
	\includegraphics[width=15cm]{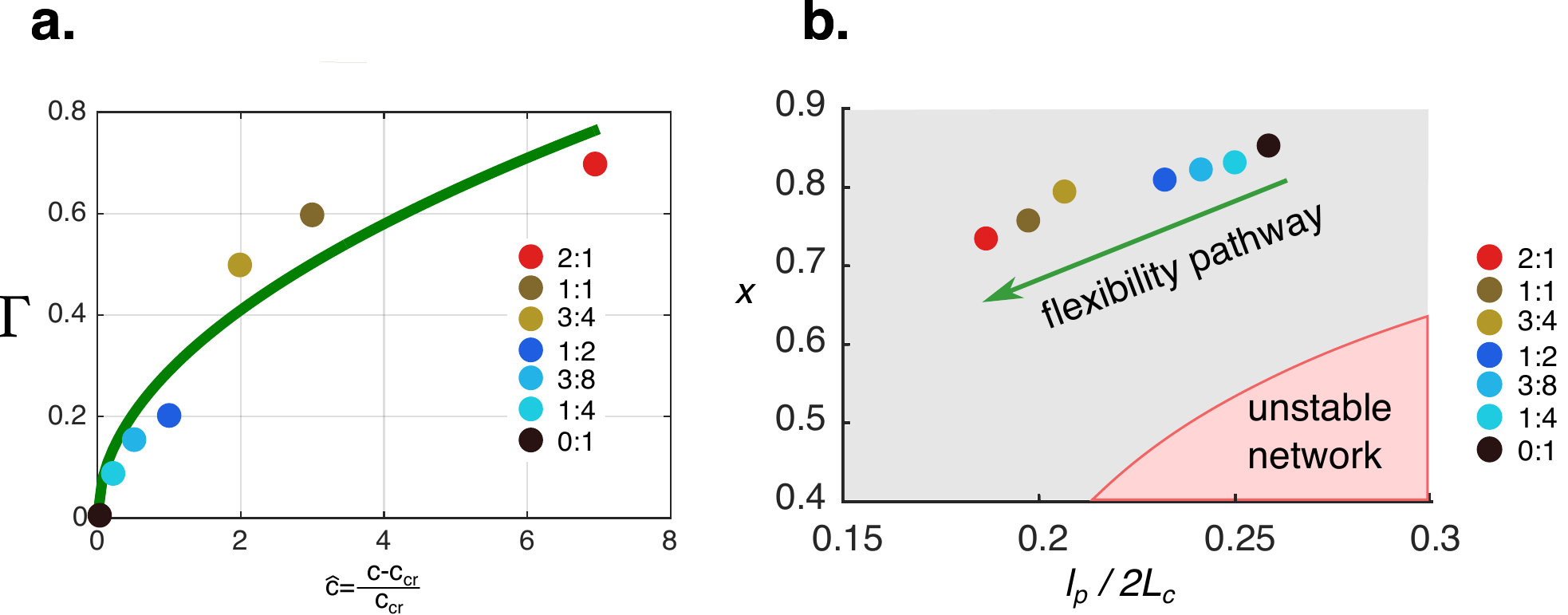}
	\centering
	\caption{\textbf{a.} Phase transition diagram showing on the horizontal axis the reduced concentration $\hat{c}$, and on the vertical axis the interaction parameter $\Gamma$. The continuous line in green describes the theoretical trend followed by the interaction parameter scales as $\Gamma\propto \hat{c}^{1/2}$, which match the path followed by the values. \textbf{b.} Stability diagram showing the distribution of the dots for different values of $c$ . The horizontal axes describes the non-dimensional stiffness, as much as the structure becomes flexible, the value will decease. On the vertical axes, the variable $x$ is proportional to the the level of strain.} \label{phased_diag}
\end{figure}
 
We found that coupling between actin and calponin encodes in a way that the emergent mechanical response at large deformations can be explained qualitatively as a second order phase transition. As can be described in the figure  \ref{phased_diag}.a, where the trend followed by the interaction parameter, $\Gamma$ as a function of the reduced concentration ratio, $\hat{c}$ is adequately described as $\Gamma\propto \hat{c}^{1/2}$, where the critical ratio is approximately $c\sim0.2$. As a whole, this suggests that the emergent process of compliance regulation via the calponin, can be condensed with arguments of phase transitions to describe the non-linear mechanics of the network at large deformations.

 \begin{figure}[h!]
	\includegraphics[width=10cm]{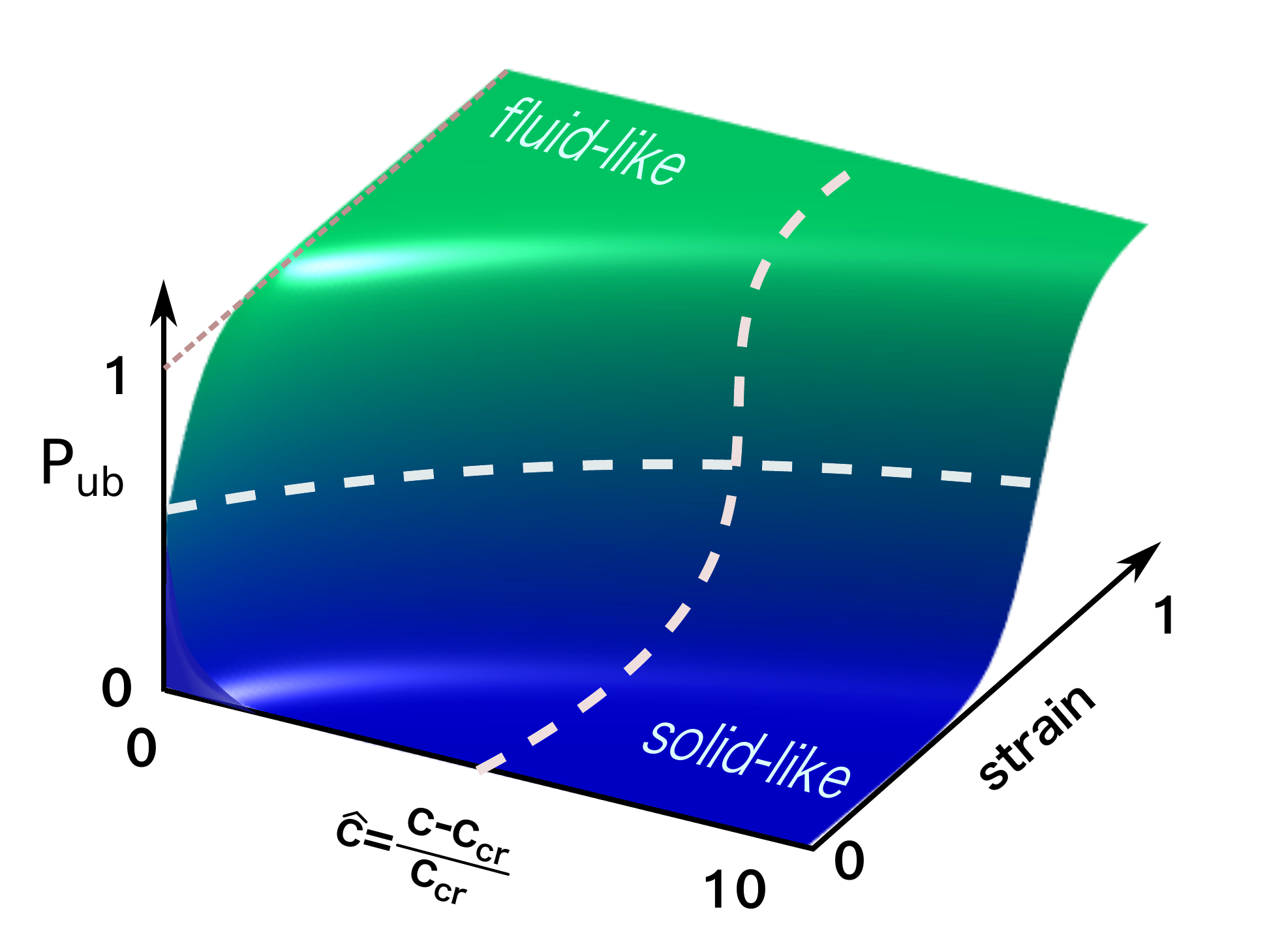}
	\centering
	\caption{Unbinding probability as a function of the concentration of calponin and the shear strain applied over the network.} \label{Pub_PHT}
\end{figure}

The figure \ref{phased_diag}.b displays the phase diagram of the stiffens-tension parameter space $(l_p/2L_c,x)$. This diagram is generated by the points that fulfil the condition of zero stress at zero strain, which satisfies the equation $\ref{refconf}$. In this way the boundary separating the positive region with $\gamma>0$ and $\sigma>0$ with the region with $\gamma<0$ and $\sigma<0$. If the network is under a pre-strain $(\epsilon>0)$ then the condition at $\gamma=0$ will leave the boundary defined by the red region towards a more stable condition, and reducing the compliance. Therefore, considering the effect introduced by the calponin over the actin network, from the perspective of the stability diagram, we can say that as the much as level of tension and stiffness decreases (increase the compliance), the ratio \textit{c=[calponin/actin]} increases. This can be seen into the figure by the general trend depicted by the green arrow.
Finally, the figure \ref{Pub_PHT} condense over the function of the unbinding probability the coupling of the effect associated with the calponin concentration scaling with a 1/2 exponent over the $\gamma_0$ that sets the transition to fluidisation.
\section*{Discussion and Conclusions}

In this work we verify, by using a mesoscale model, that the effect of calponin over the crosslinked F-actin network structure, introduces a reduction of the bundles persistence length, as was described by \cite{jensen2014}. But also a reduction in the network pre-strain which encodes as well an increment of the characteristic strain $\gamma_0$. This promotes an increment of the adhesion energy stabilising the network and allowing the mechanical structure to increase its ability to sustain higher mechanical loads before failure by extending the yielding point.

The regulation of force and failure into the gels is a very active research field where many efforts have been made to mimic biological active gels with important applications on tissue engineering or soft robotics \cite{de2019,matsuda2019}. Nevertheless, the devices to control the failure on hydrogels do not tackle that problems by using a similar mechanism as the developed by the calponin. It suggests a potential source of inspiration for future design of polymeric structures.

The presented model introduces a phase-field approach, which provides a simple description to introduce the remodelling effect exerted by calponin. As a whole, the relevance of this sort of structures motivated during the last years the development of models based on scaling laws or computational models of great complexity based on finite elements or method of particles. However, we consider that there is still place to develop mesoscale models based on mechanics of the non-linear continuum mechanics jointly with second order phase transitions. These models allow to capture the mechanics of the problem, as well as  to evaluate the plausibility of certain hypotheses without the need of complex simulations. In addition, improve the characterisation at large-deformations, which helps the definition of metrics based on the nonlinear elasticity inherent to the mechanical response. 

In summary, the proposed model provides arguments to describe the changes observed in the flexibility of the actin bundles which encodes the effects at network scale that drives the increment of adhesion and the whole stabilisation of the structure. Moreover, surprisingly the trend followed by the experiments fall with  the interaction parameter $\Gamma\sim (c-c_{cr})^{1/2}$, with the same exponent as many others physical process. Finally, the gain of a deep understanding of how the chemical pathways, as the interaction actin-calponin, interact with the mechanical structure will allow the control of their response as an allosteric material \cite{yan2017}. Will also help to clarify several biophysical processes such as mechano-transduction, with strong implications on the cellular adaptation to its environment or cell motility among others.


\begin{thebibliography}{10}

\bibitem{Schmoller2010}
K.~Schmoller, P.~Fernandez, R.~Arevalo, D.~Blair, and A.~Bausch.
\newblock {Cyclic hardening in bundled actin networks}.
\newblock {\em Nature Communications}, 1:134, 2010.

\bibitem{Schmoller2013}
K.~Schmoller and A.~Bausch.
\newblock Similar nonlinear mechanical responses in hard and soft materials.
\newblock {\em Nature materials}, 12(4):278--281, 2013.

\bibitem{lopez2016}
Horacio L{\'o}pez-Men{\'e}ndez and Jos{\'e}~F{\'e}lix Rodr{\'\i}guez.
\newblock Microstructural model for cyclic hardening in f-actin networks
  crosslinked by $\alpha$-actinin.
\newblock {\em Journal of the Mechanics and Physics of Solids}, 91:28--39,
  2016.

\bibitem{rigato2017}
Annafrancesca Rigato, Atsushi Miyagi, Simon Scheuring, and Felix Rico.
\newblock High-frequency microrheology reveals cytoskeleton dynamics in living
  cells.
\newblock {\em Nature physics}, 13(8):771, 2017.

\bibitem{lieleg2010}
Oliver Lieleg, Mireille~MAE Claessens, and Andreas~R Bausch.
\newblock Structure and dynamics of cross-linked actin networks.
\newblock {\em Soft Matter}, 6(2):218--225, 2010.

\bibitem{koll2011}
Philip Kollmannsberger and Ben Fabry.
\newblock Linear and nonlinear rheology of living cells.
\newblock {\em Annual review of materials research}, 41:75--97, 2011.

\bibitem{broedersz2014}
Chase~P Broedersz and Fred~C MacKintosh.
\newblock Modeling semiflexible polymer networks.
\newblock {\em Reviews of Modern Physics}, 86(3):995, 2014.

\bibitem{gardel2008}
Margaret~L Gardel, Karen~E Kasza, Clifford~P Brangwynne, Jiayu Liu, and David~A
  Weitz.
\newblock Mechanical response of cytoskeletal networks.
\newblock {\em Methods in cell biology}, 89:487--519, 2008.

\bibitem{fletcher2010}
Daniel~A Fletcher and R~Dyche Mullins.
\newblock Cell mechanics and the cytoskeleton.
\newblock {\em Nature}, 463(7280):485, 2010.

\bibitem{blanchoin2014}
Laurent Blanchoin, Rajaa Boujemaa-Paterski, C{\'e}cile Sykes, and Julie
  Plastino.
\newblock Actin dynamics, architecture, and mechanics in cell motility.
\newblock {\em Physiological reviews}, 94(1):235--263, 2014.

\bibitem{dos2003}
CG~Dos~Remedios, D~Chhabra, M~Kekic, IV~Dedova, M~Tsubakihara, DA~Berry, and
  NJ~Nosworthy.
\newblock Actin binding proteins: regulation of cytoskeletal microfilaments.
\newblock {\em Physiological reviews}, 83(2):433--473, 2003.

\bibitem{tseng2005}
Yiider Tseng, Thomas~P Kole, Jerry~SH Lee, Elena Fedorov, Steven~C Almo,
  Benjamin~W Schafer, and Denis Wirtz.
\newblock How actin crosslinking and bundling proteins cooperate to generate an
  enhanced cell mechanical response.
\newblock {\em Biochemical and biophysical research communications},
  334(1):183--192, 2005.

\bibitem{latorre2018}
Ernest Latorre, Sohan Kale, Laura Casares, Manuel G{\'o}mez-Gonz{\'a}lez,
  Marina Uroz, L{\'e}o Valon, Roshna~V Nair, Elena Garreta, Nuria Montserrat,
  Ar{\'a}nzazu del Campo, et~al.
\newblock Active superelasticity in three-dimensional epithelia of controlled
  shape.
\newblock {\em Nature}, 563(7730):203, 2018.

\bibitem{sun2012}
Jeong-Yun Sun, Xuanhe Zhao, Widusha~RK Illeperuma, Ovijit Chaudhuri, Kyu~Hwan
  Oh, David~J Mooney, Joost~J Vlassak, and Zhigang Suo.
\newblock Highly stretchable and tough hydrogels.
\newblock {\em Nature}, 489(7414):133, 2012.

\bibitem{nakajima2013}
Tasuku Nakajima, Takayuki Kurokawa, Saika Ahmed, Wen-li Wu, and Jian~Ping Gong.
\newblock Characterization of internal fracture process of double network
  hydrogels under uniaxial elongation.
\newblock {\em Soft Matter}, 9(6):1955--1966, 2013.

\bibitem{zhao2014}
Xuanhe Zhao.
\newblock Multi-scale multi-mechanism design of tough hydrogels: building
  dissipation into stretchy networks.
\newblock {\em Soft Matter}, 10(5):672--687, 2014.

\bibitem{ducrot2014}
E.~Ducrot, Y.~Chen, M.~Bulters, R.~Sijbesma, and C.~Creton.
\newblock Toughening elastomers with sacrificial bonds and watching them break.
\newblock {\em Science}, 344(6180):186--189, 2014.

\bibitem{suo2009}
Zhigang Suo and Jian Zhu.
\newblock Dielectric elastomers of interpenetrating networks.
\newblock {\em Applied Physics Letters}, 95(23):232909, 2009.

\bibitem{zhao2012}
Xuanhe Zhao.
\newblock A theory for large deformation and damage of interpenetrating polymer
  networks.
\newblock {\em Journal of the Mechanics and Physics of Solids}, 60(2):319--332,
  2012.

\bibitem{winder1990}
SJ~Winder and MP~Walsh.
\newblock Smooth muscle calponin. inhibition of actomyosin mgatpase and
  regulation by phosphorylation.
\newblock {\em Journal of Biological Chemistry}, 265(17):10148--10155, 1990.

\bibitem{jensen2014}
Mikkel~Herholdt Jensen, Eliza~J Morris, Cynthia~M Gallant, Kathleen~G Morgan,
  David~A Weitz, and Jeffrey~R Moore.
\newblock Mechanism of calponin stabilization of cross-linked actin networks.
\newblock {\em Biophysical journal}, 106(4):793--800, 2014.

\bibitem{meng2017}
Fanlong Meng and Eugene~M Terentjev.
\newblock Theory of semiflexible filaments and networks.
\newblock {\em Polymers}, 9(2):52, 2017.

\bibitem{vernerey2018}
Franck~J Vernerey.
\newblock Transient response of nonlinear polymer networks: A kinetic theory.
\newblock {\em Journal of the Mechanics and Physics of Solids}, 115:230--247,
  2018.

\bibitem{ferreira2018}
JPS Ferreira, MPL Parente, and RM~Natal Jorge.
\newblock Continuum mechanical model for cross-linked actin networks with
  contractile bundles.
\newblock {\em Journal of the Mechanics and Physics of Solids}, 110:100--117,
  2018.

\bibitem{Kim2009b}
T.~Kim, W.~Hwang, H.~Lee, and R.~Kamm.
\newblock {Computational analysis of viscoelastic properties of crosslinked
  actin networks.}
\newblock {\em PLoS computational biology}, 5(7), 2009.

\bibitem{zagar2011}
Goran Zagar, Patrick~R Onck, and Erik Van~der Giessen.
\newblock Elasticity of rigidly cross-linked networks of athermal filaments.
\newblock {\em Macromolecules}, 44(17):7026--7033, 2011.

\bibitem{borau2012}
Carlos Borau, Taeyoon Kim, Tamara Bidone, Jos{\'e}~Manuel Garc{\'\i}a-Aznar,
  and Roger~D Kamm.
\newblock Dynamic mechanisms of cell rigidity sensing: insights from a
  computational model of actomyosin networks.
\newblock {\em PLoS One}, 7(11):e49174, 2012.

\bibitem{vzagar2015}
Goran {\v{Z}}agar, Patrick~R Onck, and Erik van~der Giessen.
\newblock Two fundamental mechanisms govern the stiffening of cross-linked
  networks.
\newblock {\em Biophysical journal}, 108(6):1470--1479, 2015.

\bibitem{tito2019h}
Nicholas~B Tito, Costantino Creton, Cornelis Storm, and Wouter~G Ellenbroek.
\newblock Harnessing entropy to enhance toughness in reversibly crosslinked
  polymer networks.
\newblock {\em Soft Matter}, 2019.

\bibitem{blundell2009}
JR~Blundell and EM~Terentjev.
\newblock Stretching semiflexible filaments and their networks.
\newblock {\em Macromolecules}, 42(14):5388--5394, 2009.

\bibitem{meng2016}
Fanlong Meng and Eugene~M Terentjev.
\newblock Nonlinear elasticity of semiflexible filament networks.
\newblock {\em Soft Matter}, 12(32):6749--6756, 2016.

\bibitem{lopez2017}
Horacio Lopez-Menendez and Jose~Felix Rodriguez.
\newblock Towards the understanding of cytoskeleton
  fluidisation--solidification regulation.
\newblock {\em Biomechanics and modeling in mechanobiology}, 16(4):1159--1169,
  2017.

\bibitem{lopez2019}
Horacio Lopez-Menendez and Joseph D'Alessandro.
\newblock Unjamming and nematic flocks in endothelial monolayers during
  angiogenesis: theoretical and experimental analysis.
\newblock {\em Journal of the Mechanics and Physics of Solids}, 125:74--88,
  2019.

\bibitem{nishimori2010}
Hidetoshi Nishimori and Gerardo Ortiz.
\newblock {\em Elements of phase transitions and critical phenomena}.
\newblock OUP Oxford, 2010.

\bibitem{Arruda1993}
E.~Arruda and M.~Boyce.
\newblock {A three-dimensional constitutive model for the large stretch
  behaviour of rubber elastic materials}.
\newblock {\em Journal of the Mechanics and Physics of Solids}, 41:389--412,
  1993.

\bibitem{Palmer2008}
J.~Palmer and M.~Boyce.
\newblock {Constitutive modeling of the stress-strain behavior of F-actin
  filament networks.}
\newblock {\em Acta biomaterialia}, 4(3):597--612, 2008.

\bibitem{ferrer2008}
Jorge~M Ferrer, Hyungsuk Lee, Jiong Chen, Benjamin Pelz, Fumihiko Nakamura,
  Roger~D Kamm, and Matthew~J Lang.
\newblock Measuring molecular rupture forces between single actin filaments and
  actin-binding proteins.
\newblock {\em Proceedings of the National Academy of Sciences},
  105(27):9221--9226, 2008.

\bibitem{de1979}
Pierre-Gilles De~Gennes.
\newblock {\em Scaling concepts in polymer physics}.
\newblock Cornell university press, 1979.

\bibitem{Brown2009}
A.~Brown, R.~Litvinov, D.~Discher, P.~Purohit, and J.~Weisel.
\newblock {Multiscale mechanics of fibrin polymer: gel stretching with protein
  unfolding and loss of water.}
\newblock {\em Science}, 325(5941):741--4, 2009.

\bibitem{purohit2011}
P~Purohit, R~Litvinov, A~Brown, D~Discher, and J~Weisel.
\newblock Protein unfolding accounts for the unusual mechanical behavior of
  fibrin networks.
\newblock {\em Acta biomaterialia}, 7(6):2374--2383, 2011.

\bibitem{Bell1978}
G.~Bell.
\newblock {Models for the specific adhesion of cells to cells}.
\newblock {\em Science}, 200(4342):618--627, 1978.

\bibitem{Golji2009}
J.~Golji, R.~Collins, and M.~Mofrad.
\newblock Molecular mechanics of the $\alpha$-actinin rod domain: Bending,
  torsional, and extensional behavior.
\newblock {\em PLoS computational biolog}, 5, 2009.

\bibitem{de2019}
Paula de~Almeida, Maarten Jaspers, Sarah Vaessen, Oya Tagit, Giuseppe Portale,
  Alan~E Rowan, and Paul~HJ Kouwer.
\newblock Cytoskeletal stiffening in synthetic hydrogels.
\newblock {\em Nature communications}, 10(1):609, 2019.

\bibitem{matsuda2019}
Takahiro Matsuda, Runa Kawakami, Ryo Namba, Tasuku Nakajima, and Jian~Ping
  Gong.
\newblock Mechanoresponsive self-growing hydrogels inspired by muscle training.
\newblock {\em Science}, 363(6426):504--508, 2019.

\bibitem{yan2017}
Le~Yan, Riccardo Ravasio, Carolina Brito, and Matthieu Wyart.
\newblock Architecture and coevolution of allosteric materials.
\newblock {\em Proceedings of the National Academy of Sciences},
  114(10):2526--2531, 2017.

\end{thebibliography}
\end{document}